\documentstyle[11pt,paspconf]{article}

\def\etal{\it et al. \rm}

\def\gsim{ \lower .75ex \hbox{$\sim$} \llap{\raise .27ex \hbox{$>$}} }
\def\lsim{ \lower .75ex \hbox{$\sim$} \llap{\raise .27ex \hbox{$<$}} }
\def\pp{\noindent\parshape 2 0truecm 16truecm 1truecm 15truecm}

\def\mn{{\it M.N.R.A.S, }}
\def\apj{{\it Ap.J, }}
\def\aj{{\it A.J., }}

\def\nat{{\it Nature, }}

\def\spose#1{\hbox to 0pt{#1\hss}}
\def\simlt{\mathrel{\spose{\lower 3pt\hbox{$\mathchar"218$}}
     \raise 2.0pt\hbox{$\mathchar"13C$}}}
\def\simgt{\mathrel{\spose{\lower 3pt\hbox{$\mathchar"218$}}
     \raise 2.0pt\hbox{$\mathchar"13E$}}}
\def\gsim{ \lower .75ex \hbox{$\sim$} \llap{\raise .27ex \hbox{$>$}} }
\def\lsim{ \lower .75ex \hbox{$\sim$} \llap{\raise .27ex \hbox{$<$}} }
\def\simprop{ \lower .75ex \hbox{$\sim$} \llap{\raise .27ex \hbox{$\propto$}} }

\begin{document}

\newread\epsffilein    
\newif\ifepsffileok    
\newif\ifepsfbbfound   
\newif\ifepsfverbose   
\newdimen\epsfxsize    
\newdimen\epsfysize    
\newdimen\epsftsize    
\newdimen\epsfrsize    
\newdimen\epsftmp      
\newdimen\pspoints     
\pspoints=1bp          
\epsfxsize=0pt         
\epsfysize=0pt         
\def\epsfbox#1{\global\def\epsfllx{72}\global\def\epsflly{72}%
   \global\def\epsfurx{540}\global\def\epsfury{720}%
   \def\lbracket{[}\def\testit{#1}\ifx\testit\lbracket
   \let\next=\epsfgetlitbb\else\let\next=\epsfnormal\fi\next{#1}}%
\def\epsfgetlitbb#1#2 #3 #4 #5]#6{\epsfgrab #2 #3 #4 #5 .\\%
   \epsfsetgraph{#6}}%
\def\epsfnormal#1{\epsfgetbb{#1}\epsfsetgraph{#1}}%
\def\epsfgetbb#1{%
%
%
\openin\epsffilein=#1
\ifeof\epsffilein\errmessage{I couldn't open #1, will ignore it}\else
%
%
   {\epsffileoktrue \chardef\other=12
    \def\do##1{\catcode`##1=\other}\dospecials \catcode`\ =10
    \loop
       \read\epsffilein to \epsffileline
       \ifeof\epsffilein\epsffileokfalse\else
%
%
          \expandafter\epsfaux\epsffileline:. \\%
       \fi
   \ifepsffileok\repeat
   \ifepsfbbfound\else
    \ifepsfverbose\message{No bounding box comment in #1; using defaults}\fi\fi
   }\closein\epsffilein\fi}%
%
%
\def\epsfsetgraph#1{%
   \epsfrsize=\epsfury\pspoints
   \advance\epsfrsize by-\epsflly\pspoints
   \epsftsize=\epsfurx\pspoints
   \advance\epsftsize by-\epsfllx\pspoints
%
%
   \epsfxsize\epsfsize\epsftsize\epsfrsize
   \ifnum\epsfxsize=0 \ifnum\epsfysize=0
      \epsfxsize=\epsftsize \epsfysize=\epsfrsize
%
%
     \else\epsftmp=\epsftsize \divide\epsftmp\epsfrsize
       \epsfxsize=\epsfysize \multiply\epsfxsize\epsftmp
       \multiply\epsftmp\epsfrsize \advance\epsftsize-\epsftmp
       \epsftmp=\epsfysize
       \loop \advance\epsftsize\epsftsize \divide\epsftmp 2
       \ifnum\epsftmp>0
          \ifnum\epsftsize<\epsfrsize\else
             \advance\epsftsize-\epsfrsize \advance\epsfxsize\epsftmp \fi
       \repeat
     \fi
   \else\epsftmp=\epsfrsize \divide\epsftmp\epsftsize
     \epsfysize=\epsfxsize \multiply\epsfysize\epsftmp   
     \multiply\epsftmp\epsftsize \advance\epsfrsize-\epsftmp
     \epsftmp=\epsfxsize
     \loop \advance\epsfrsize\epsfrsize \divide\epsftmp 2
     \ifnum\epsftmp>0
        \ifnum\epsfrsize<\epsftsize\else
           \advance\epsfrsize-\epsftsize \advance\epsfysize\epsftmp \fi
     \repeat     
   \fi
%
%
   \ifepsfverbose\message{#1: width=\the\epsfxsize, height=\the\epsfysize}\fi
   \epsftmp=10\epsfxsize \divide\epsftmp\pspoints
   \newcount\figskipcount
      \message{#1 \the\epsfysize  }
   \vbox to\epsfysize{\vfil\hbox to\epsfxsize{%
      \includegraphics{#1}%
      \hfil}}%
\epsfxsize=0pt\epsfysize=0pt}%

%
%
{\catcode`\%=12 \global\let\epsfpercent=
%
%
\long\def\epsfaux#1#2:#3\\{\ifx#1\epsfpercent
   \def\testit{#2}\ifx\testit\epsfbblit
      \epsfgrab #3 . . . \\%
      \epsffileokfalse
      \global\epsfbbfoundtrue
   \fi\else\ifx#1\par\else\epsffileokfalse\fi\fi}%
%
%
\def\epsfgrab #1 #2 #3 #4 #5\\{%
   \global\def\epsfllx{#1}\ifx\epsfllx\empty
      \epsfgrab #2 #3 #4 #5 .\\\else
   \global\def\epsflly{#2}%
   \global\def\epsfurx{#3}\global\def\epsfury{#4}\fi}%
%
%
\def\epsfsize#1#2{\epsfxsize}
%
%
\let\epsffile=\epsfbox

\def\figinsert#1#2{\epsfbox{#1} \message{#2} }    

\title{Galaxy formation and evolution: what to expect from hierarchical
clustering models}


\author{C.S. Frenk, C.M. Baugh and S. Cole}
\affil{Physics Department, University of Durham, UK}


\begin{abstract}
We give a brief review of current theoretical work in galaxy
formation. Recent results from N-body and N-body/hydrodynamic simulations, 
and from semianalytic modelling are discussed.  We present updated versions
of some figures from Cole et al (1994).  In particular, we show the effect
of using the revised stellar population synthesis model of Bruzual and
Charlot, which results in a much better match to the observed colour 
distribution of galaxies than before. We also compare the model output 
with recently available data on the galaxy luminosity function and the
redshift distribution of galaxies in the B and K bands.
The form of the Tully-Fisher relation at high redshift predicted by our 
semi-analytic scheme for galaxy formation is given.
\end{abstract}

\section{Introduction}

In hierarchical clustering theories of galaxy formation, galaxies form by
gas cooling and condensing into dark matter halos which, in turn, form by a
hierarchy of mergers (White \& Rees 1978). The context in which this
process takes place is specified by a cosmological model that determines
the spectrum of primordial density fluctuations and the rate at which they
grow by gravitational instability.  The best known example of such a model
is the cold dark matter (CDM) model (see Frenk 1991 for a review), but a
number of alternatives (mostly variants of CDM), have recently become
popular in response to new data on large-scale structure and the COBE
detection of anisotropies in the microwave background radiation. Regardless
of the specific cosmological model that one wishes to consider, there are
at least six distinct physical processes that need to be included in any
theory of galaxy formation:

\begin{itemize}
\item{1.} The growth of dark matter halos by accretion and mergers.
\item{2.} The dynamics of cooling gas. 
\item{3.} Star formation.
\item{4.} Energy feedback into prestellar gas from the products of stellar 
evolution.
\item{5.} Evolution of the stellar populations that form.
\item{6.} Galaxy mergers. 
\end{itemize}

A number of theoretical tools have been developed over the years to
investigate these processes, both individually and collectively. N-body
simulations have led to significant progress in understanding process {\it
(1.)}, while the recently developed N-body/hydrodynamic techniques are
beginning to address processes {\it (1-4)} and {\it (6)}. In addition,
semianalytic modelling, a relatively new tool, can treat all six processes
together and thus explore the effects of different assumptions on the
properties of the galaxy population as a whole. In this review, we will
outline some of the areas where progress has been made and highlight some
as yet unresolved issues.

\section{ Physical processes} 

\subsection{Evolution of dark matter halos} 

The main features of the formation of dark matter halos by hierarchical
clustering were already established in N-body simulations carried out
a decade ago (eg. Frenk \etal 1985, 1988; Efstathiou \etal 1988). A
protohalo perturbation, initially expanding at a reduced rate, collapses,
often into filamentary or sheet-like structures, which subsequently break
up into roughly spherical lumps. These merge together producing a centrally
concentrated and essentially smooth dark halo.  This process is illustrated in
Figure~1 which shows the development of a galactic halo in a flat
`low'-$\Omega$ CDM model.

One of the main early results from N-body simulations was the realisation
that the rotation curves of dark galactic halos in the standard CDM
model are approximately flat, suggesting an explanation for the inferred
structure of the halos of spiral galaxies (Frenk \etal 1985, Quinn \etal
1986). These simulations, however, were limited in particle number and did
not resolve the inner regions of the halos where the visible galaxy
actually forms. This issue has recently been addressed in a series of
high-resolution simulations by Navarro, Frenk \& White (1995).
 The density
profiles of galactic halos in the CDM model show noticeable departures from
an $r^{-2}$ law, gently sloping from $r^{-1}$ near the centre to
$r^{-3}$ near the virial radius. When the gravitational effect of a
disk is included, the resulting rotation
curves agree well with observations of galaxies, from dwarfs to bright
galaxies, provided the disks fulfill two conditions: (i) their
stellar mass-to-light ratio increases roughly as $L^{1/2}$ and (ii) the baryon
fraction increases with luminosity such that for galaxies with observed
circular velocity, $V_c \gsim 200 {\rm km s}^{-1}$, there is only a weak of correlation
between this velocity and total halo mass. It is unclear whether the
observed disks of spirals satisfy these conditions.

A second important early result concerns the angular momentum of galactic
halos. This is acquired through tidal torques and, in the
linear regime grows linearly with time (eg White 1995). Tidal effects during
merging events efficiently transfer the angular momentum invested
in the orbits of the merging subclumps into the outer halo and, as a
result, the inner parts of merger remnants end up rotating slowly (Frenk
\etal 1985, Barnes \& Efstathiou 1987, Cole \& Lacey
1995). This non-linear feature has often been invoked as a possible
explanation for the low rotation speeds of elliptical galaxies.

\begin{figure}
\vspace{15cm}  
\caption{The formation of a galactic dark matter halo in an N-body
simulation. The left-hand column shows the projected particle distribution, in
comoving coordinates, of a cubical region of present comoving length 27
Mpc. The right hand column shows, now in physical coordinates, the growth of
the large clump seen in the bottom left of the region. Each panel on the
right hand row has length 3.8 Mpc. From top to bottom the epochs shown
correspond to $z=5,0.5$ and 0. The parameters of the simulation are:
mean density, $\Omega=0.3$; cosmological constant, $\Lambda=0.7$; Hubble
constant, $H_0=100 h$km~s$^{-1}$~Mpc$^{-1}$, with $h=0.7$; and spectrum
normalisation, $\sigma_8=1.14$ (as inferred from the COBE data).  The
simulation followed 262144 particles and was performed the T3D parallel
supercomputer at Edinburgh.
}
\end{figure}

\subsection{ The dynamics of cooling gas} 

The main ideas here were put forward nearly twenty years ago by
Rees \& Ostriker (1977), Silk (1977) and Binney (1977). When a dark matter
halo collapses, any gas admixed with it will also collapse, but whereas the
dark matter free streams, the gas is shock heated. These early
papers assumed that shocks would heat up the gas to the virial temperature
of the halo, an assumption verified -- in the non-radiative limit -- in the
N-body/hydrodynamic simulations of Evrard (1990).  These and subsequent
simulations (eg. Katz \& White 1993, Navarro, Frenk \&
White 1995) also showed that, in this limit, the gas acquires a density
profile that closely parallels that of the dark matter.
Rees \& Ostriker argued that if 
the cooling time of virialised gas was shorter than its 
dynamical time, the gas would
collapse to make a galaxy.
White \& Rees (1978) recognized, however, that in a hierarchical
model this simple scheme would lead to a cooling catastrophe since at
early times the density is so high that all the gas would cool into
subgalactic lumps where it would presumably turn into stars.
This patently did not happen in the
universe - there is still plenty of gas
around today. 
White \& Rees solved this problem by introducing the idea
of feedback, whereby the energy released by supernovae associated with
an early generation of stars reheats some of the gas before it has had
a chance to condense into halos at high redshift. (Efstathiou 1992 has
argued that photoionisation by a UV background at high redshift would
have a similar effect.)

Testing these simple physical arguments in numerical simulations is
difficult because the propensity of the gas to cool at high density implies
that the behaviour of the gas is always determined by the resolution limit
of the simulation. This numerical artefact, however, can be turned to
advantage if it is loosely interpreted as an effective source of
feedback. N-body/hydrodynamic simulations of representative cosmological
volumes in which the gas is allowed to cool are still at an early stage (eg
Katz \etal 1992, Cen \& Ostriker 1992, Frenk \etal 1995) and show
that the behaviour of the gas is more complex than expected in the simple
analytic picture.
Simulations of the formation of individual galaxies
produce disks, often with beautiful spiral arms (e.g.
Steinmetz \& Muller 1995), but these disks rotate much too slowly. This is
because merger events transfer angular momentum from gas fragments to the
outer dark matter halo in much the same way as the mergers of collionsless
particles do (Navarro, Frenk \& White 1995).
This angular momentum problem
for disks remains a major unresolved issue in studies of galaxy
formation. One possible solution may be, again, to invoke some form of
feedback which might keep the gas hot and allow it to cool slowly rather to
be collected in subclumps.

\subsection{ Star formation and feedback} 

Current understanding of star formation and the attendant feedback, in the
context of galaxy formation, is laughably poor. All that can be done at
present is to try and model these processes in a heuristic fashion. For
example, in an N-body/hydrodynamic simulation one can stipulate a number of
conditions for gas to turn into stars, eg, that it be cool and dense (ie
above the Jeans mass) and that it be inflowing into a halo.
Systematic tests of such algorithms are just beginning 
(e.g. Navarro \& White 1993).

\subsection{ Galaxy mergers} 

Simulations of the merging of individual galaxy pairs or small groups have
a long and distinguished history (see for example Barnes (1996)). 
Such simulations address issues such as the structure and
rotation properties of merger remnants, or the gas flows triggered by
mergers. From the point of view of galaxy formation in general, a key issue
is the relative timescale for the merging of dark matter halos and the
galaxies they harbour. As a consequence of their higher binding energy,
galaxies take longer to merge than their halos. 
Furthermore, the simulations show that galaxies (or at any rate the
clumps of cool gas identified with galaxies in the models) merge on a
dynamical friction timescale, provided that the mass that is input into
Chandrasekhar's classic formula is the total, gas plus dark matter, mass of
the merging satellite (Navarro, Frenk \& White 1995).

\section{Semianalytic models}

\begin{figure}
\begin{picture}(100, 350)(0,0)
\put(-10,-8)
{\epsfxsize=13.truecm \epsfysize=13.truecm 
\epsfbox[25 200 550 750]{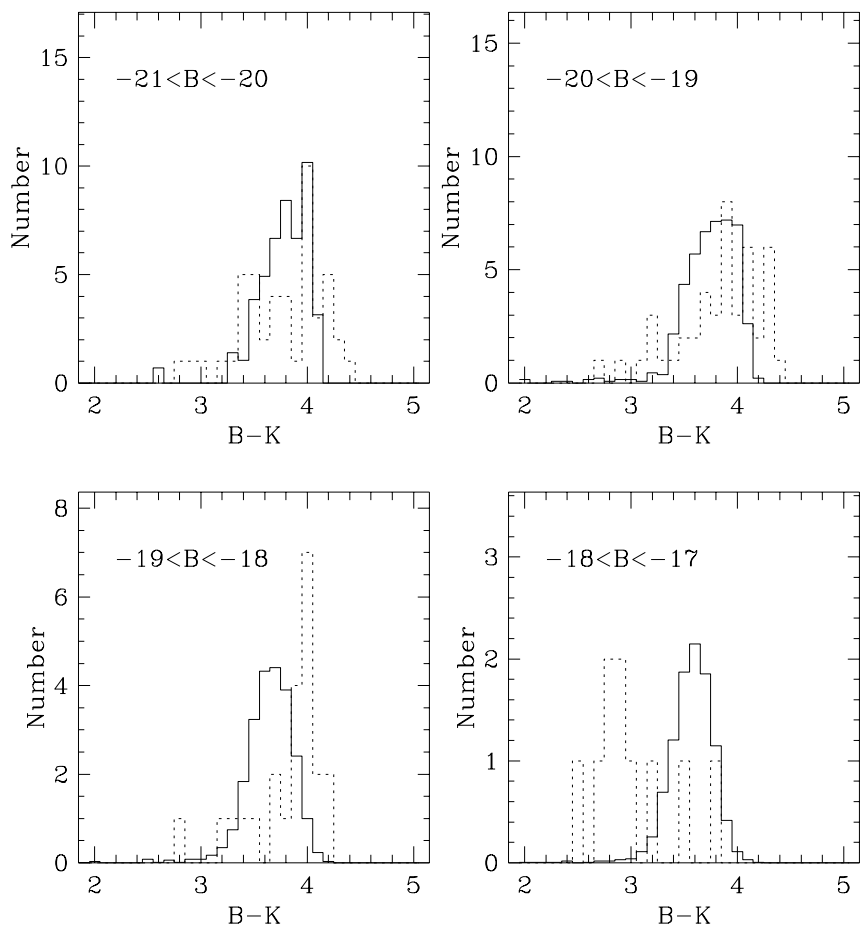}}
\end{picture}
\caption[junk]{Histograms of B-K colour distributions for various 
ranges of B absolute magnitude. The broken lines are data from 
Mobasher {\it et al.} (1986) and show the observed number of galaxies in 
the data set. 
The model output is shown by the solid lines which have been normalised 
to enclose the same area as the data.
}
\label{fig:col}
\end{figure}

\begin{figure}
\centering
\centerline
{\epsfxsize=13.5truecm \epsfysize=13.5truecm 
\epsfbox{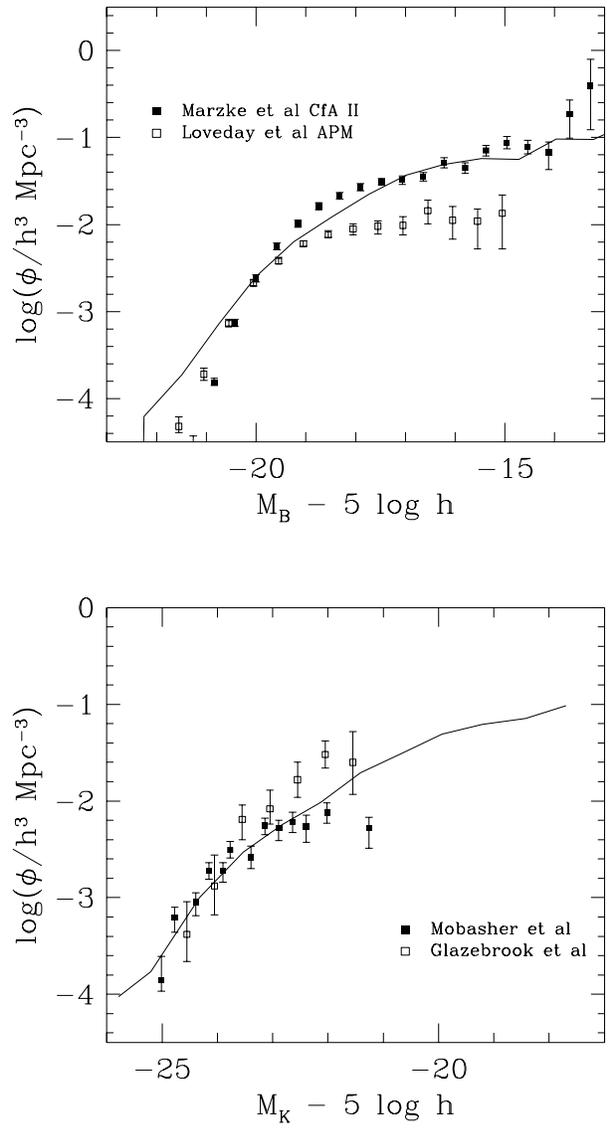}}
\caption[junk]{The luminosity function of our model (solid line)
compared with the B-band data of Loveday \etal (1992) and Marzke \etal 
(1994).We have normalised the model to match the 
knee of the B-band luminosity function.
The lower panel shows the comparison with the K-band data of 
Mobasher \etal (1986) and Glazebrook \etal (1994).
\label{fig:lf}}
\end{figure}

\begin{figure}
\begin{picture}(100, 250)(0,0)
\put(40,10)
{\epsfxsize=10.truecm \epsfysize=10.truecm 
\epsfbox[80 430 470 740]{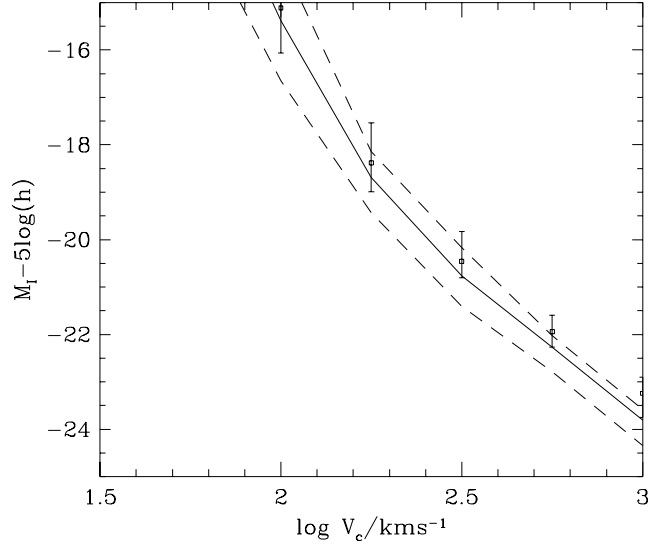}}
\end{picture}
\caption[junk]{
The Tully-Fisher relation predicted by our model as a function 
of redshift. The solid line shows the median magnitude in bins 
of circular velocity at $z=0$; the dashed lines show the 20 and $80^{th}$ 
percentiles.
The points show the median and the errorbars show the location of the 
percentiles at $z=0.5$.}
\label{fig:tf1}
\end{figure}

\begin{figure}
\begin{picture}(100, 100)(0,0)
\put(-25, 0)
{ \epsfysize = 3.8cm 
\epsfbox[65 530 300 730]{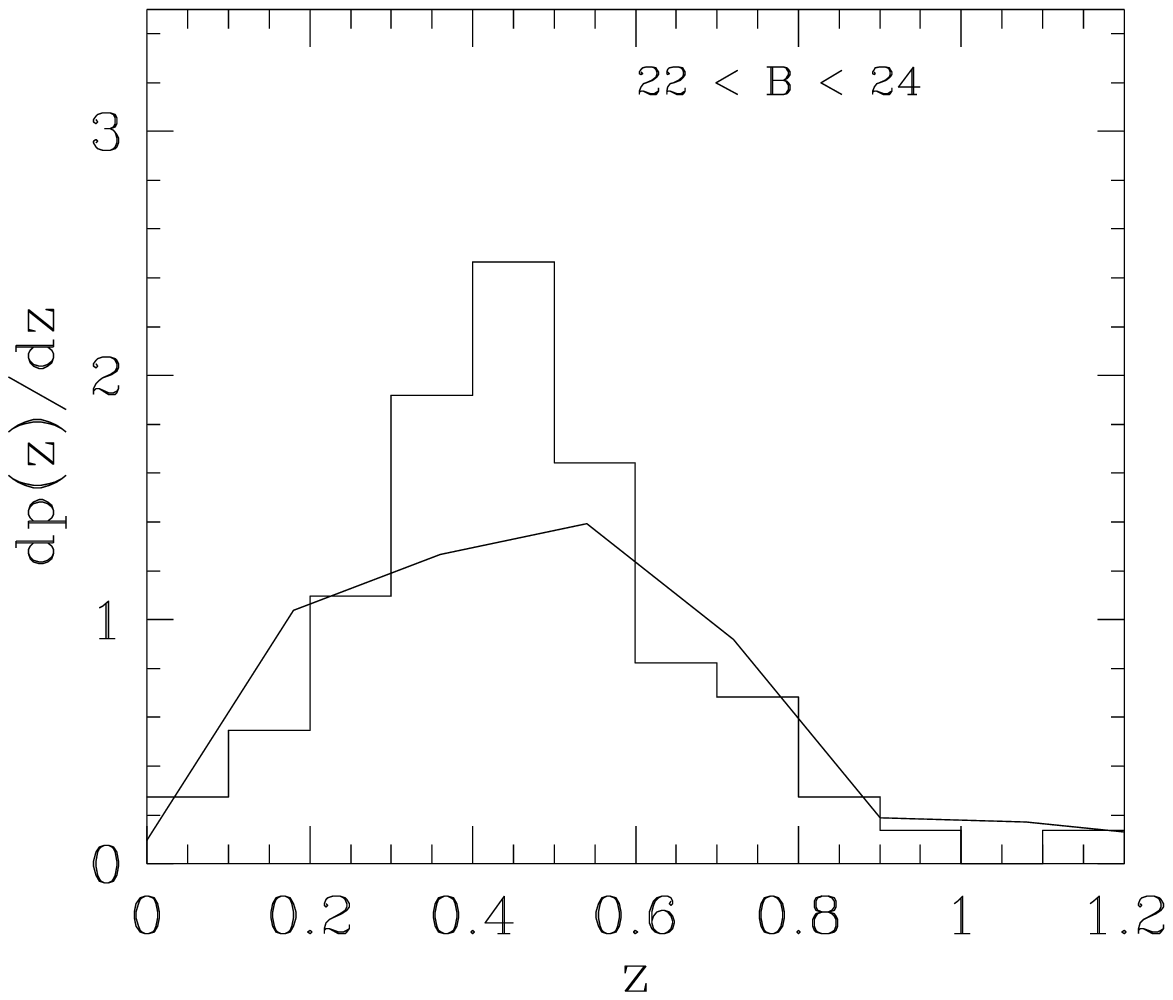}} 
\put (160, 0)
{\epsfysize = 3.8cm 
\epsfbox[65 530 300 730]{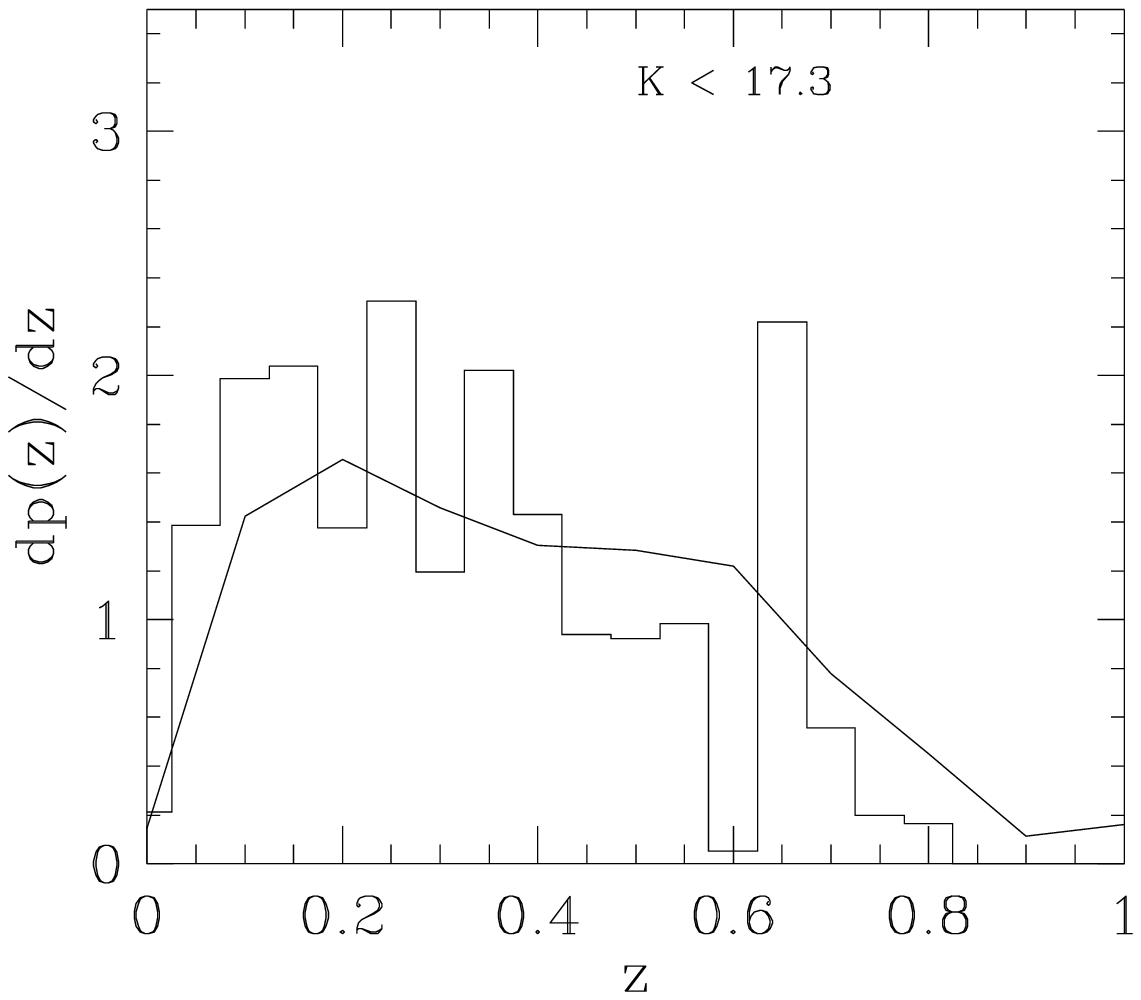} }
\end{picture}
\vspace{1.6 cm}
\caption[junk]{The redshift distribution of model galaxies compared
with new data for K and B-band selected samples of 
Glazebrook {\it et al.} 1995a,b.
The histograms show the observed distributions and the curves show 
the model galaxy redshifts.
The K band data consists of 124 redshifts for galaxies with 
$ K < 17.3$ and is weighted for incompleteness.
The B band sample is $70 \%$ complete and contains 70 galaxies 
with $22.5 < B <24$.}
\label{fig:dndz}
\end{figure}

\begin{figure}
{\epsfxsize=14.truecm \epsfysize=14.truecm 
\epsfbox{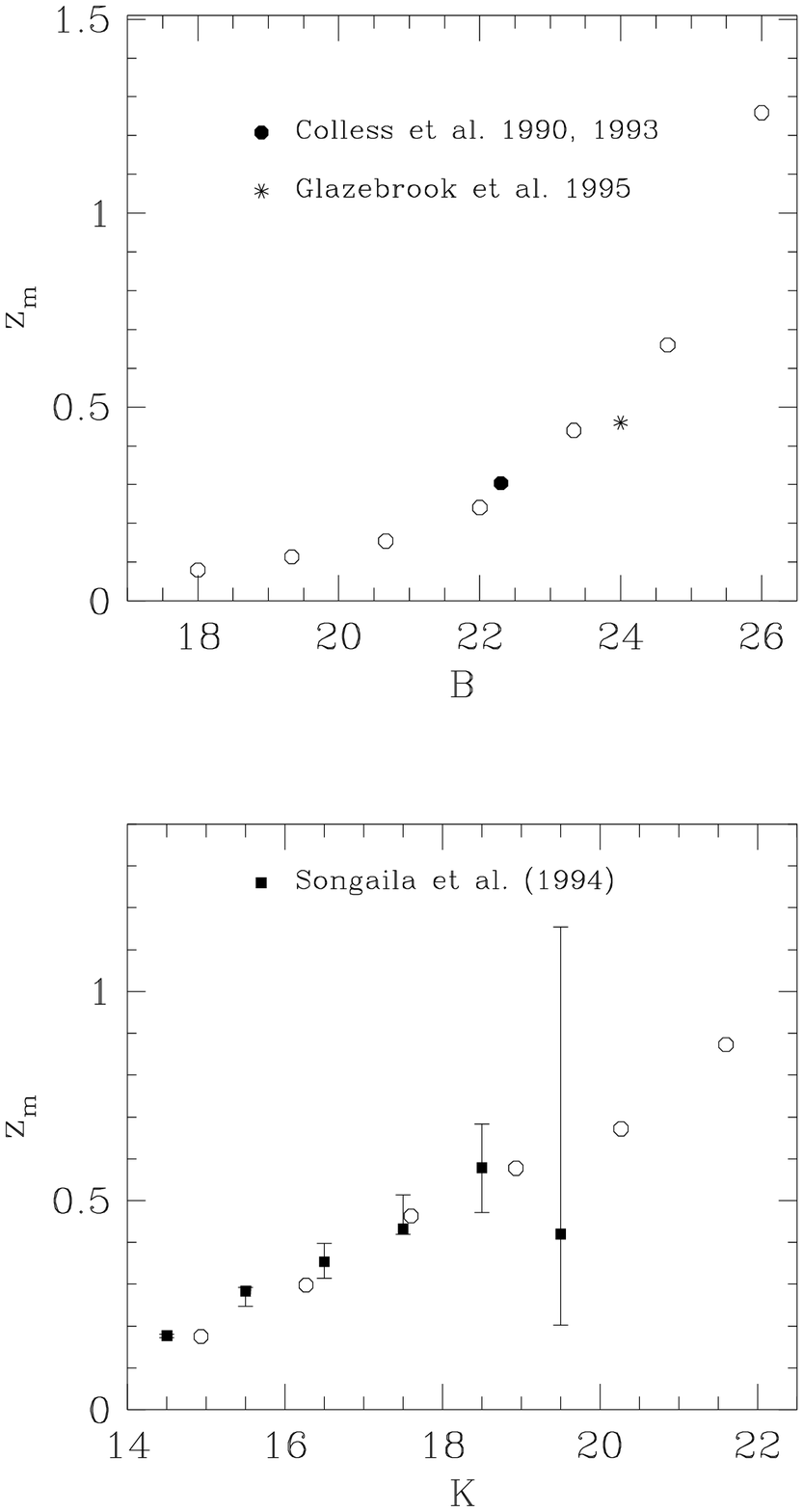}}
\caption[junk]{The median redshift as a function of limiting 
apparent magnitude in the B and K bands.
The open points show the predictions of our semi-analytic model.
\label{fig:zmed}}
\end{figure}

Our understanding of the full range of complex phenomena listed in the
Introduction can be approximated by a set of simple rules.  These
rules can then form the basis of a semianalytic model for galaxy
formation, that follows the collapse and mergers of dark matter halos
and the star formation histories of the galaxies.  Such models
(Kauffman {\it et al.} 1993, Lacey {\it et al.} 1993, Cole {\it et
al.} 1994) have been successful in accounting for the general
properties of the observed galaxy distribution, such as the shape of
the luminosity function, faint number counts and colours.  However, a
number of fundamental problems remain that appear to suggest that the
modelling of the processes (1 - 6) needs to be improved, rather than
altering the choice of cosmology (Heyl {\it et al.} 1994).

In the original model of Cole et al. (1994), the model galaxies were 
not as red as many observed ellipticals.
A study of several stellar population codes (Charlot {\it et al.} 1995) 
has led to a revision of the Bruzual and Charlot (1993) models.
This has resulted in the model galaxies being typically 0.2~mag redder
in $B-K$.
An updated comparison of the colour distribution of the model galaxies 
from Cole et al. with observed colours is given in Figure~\ref{fig:col}.

The luminosity function of Cole et al. was flatter than that achieved 
by other semianalytic models, because of the strong feedback adopted, which 
severely restricts star formation in halos of low circular velocity.
However, this model still predicts more faint galaxies than are observed 
(Loveday et al 1992), though recent results indicate that the faint end 
of the luminosity function is still uncertain (McGaugh 1994, Marzke 
{\it et al.} 1994).
A comparison of the luminosity function predicted by our model with 
the data of Loveday \etal and Marzke \etal is  given in Figure 
\ref{fig:lf}.
The lower panel shows the comparison with the K-band data of 
Mobasher \etal (1986) and Glazebrook \etal (1994).
Following Glazebrook \etal, we have corrected the Mobasher \etal 
magnitudes by +0.22 mag., to compensate for the different k 
corrections used, and we have applied a -0.3 correction to 
Glazebrook \etal's magnitudes, so that they correspond to 
the $40 h^{-1} {\rm Kpc}$ aperature used by Mobasher \etal.

The Tully-Fisher relation recovered by Cole {\it et al} gives a good match 
to the observed scatter and slope at zero redshift (see Figure 11 of 
Cole {\it et al}).
However, there is an offset between the observed relation and the 
prediction of the model, which suggests that the model galaxies are 
either too faint by about 1.5~mag or are in halos that have a circular 
velocity that is $\sim 60 \% $ too high.
This can be traced back to an overproduction in CDM-like cosmologies of halos 
typical of those that contain luminous galaxies.

Both the Tully-Fisher and luminosity function problems could be related 
to surface brightness effects. 
We plan to incorporate a scale length 
into the models, allowing us to select only those 
galaxies above some surface brightness threshold.

Figure \ref{fig:tf1} shows the evolution of the Tully-Fisher relation 
predicted by our model.
The solid line shows the median magnitude in bins of circular 
velocity at $z=0$.
The dashed lines show the location of the $20^{th}$ and $80^{th}$ 
percentiles.
The points show the position of the median magnitude as a function 
of circular velocity at redshift $z=0.5$; the errorbars here indicate the 
location of the 20 and $80^{th}$ percentiles.

We also present an updated version of the redshift distributions
predicted by the model of Cole \etal  
Figure~\ref{fig:dndz} compares
the model predictions with the recent redshift survey data
 of Glazebrook {\it et al.} (1995a,b). The model and observed
distributions are in very good agreement.
We plot the median redshift as a function of limiting apparent 
magnitude in Figure \ref{fig:zmed}.
The predictions of our model agree well with the faint redshift 
data currently available.

We have extended the model to split the light of each galaxy up into 
a bulge and a disk component.
Stars are formed in a disk when gas is accreted from 
the dark matter halo.
Bulges are formed in violent merger events, which destroy the disks 
of the progenitors and are accompanied by a burst of star formation.
This allows us to make a broad morphological classification of our 
galaxies and make predictions 
of galaxy properties as a function of bulge to disk luminosity ratio.
We set the parameters that define the strength of a merger event and 
the bulge to disk ratios that distinguish between different 
morphological types by requiring that our model 
reproduces the local morphological mix.
We are then able predict the mean colour and 
scatter in colour for different morphological classes in different 
environments, the mix of types in different environments as a 
function of redshift and the faint counts for the various galaxy types. 
We find good agreement with the faint HST counts of Glazebrook \etal 
(1995c) and recover the type of evolution in cluster membership 
reported by Butcher \& Oemler (1978) (Baugh \etal 1995).

\section*{References}

\pp Barnes, J., 1996, to appear in the proceedings of IAU 171, 
   ``New Light \\ on 
    Galaxy Evolution'' eds., Bender, R., Davies, R.L., (Kluwer)

\pp Barnes, J., Efstathiou, G., 1987, \apj 319, 575

\pp Baugh, C.M., Cole, S., Frenk, C.S., 1995, \mn submitted

\pp Binney, J.J., 1977, \apj 215, 483

\pp Bruzual, G.,  Charlot, S.,  1993, \apj 405, 538

\pp Butcher, H., Oemler, A., 1978, \apj 219, 18

\pp Cen, R., Ostriker, J.P., 1992, \apj 339, 331 

\pp Charlot, S., Worthey, G., Bressan, A., 1995 \apj in press

\pp Cole, S., Lacey, C.G., 1995 \mn submitted

\pp Cole, S., Aragon-Salamanca, A., Frenk, C.S., Navarro, J., Zepf, S.,  
1994, \\ \mn 271, 781

\pp Colless, M.M., Ellis, R.S., Taylor, K.,  Hook, R.N., 1990, \mn 244, 408

\pp Colless, M.M., Ellis, R.S., Broadhurst T., Taylor, K., Peterson, B.A., \\
    1993, \mn 261, 19

\pp Efstathiou, G. 1992, \mn 256, 43p

\pp Efstathiou, G., Frenk, C.S., White, S.D.M., Davis, M., 1988, 
\mn \\ 235, 715.

\pp Evrard, A.E. 1990, {\it Ap.J.}, 363, 349.

\pp Frenk, C.S. 1991, in {\it The Birth and Early Evolution of our 
Universe}, 
Nobel \\ Symp.  No 79, eds J.S. Nilsson, \etal, 
World Sci., {\it Physica Scripta}, T36, 70.

\pp Frenk, C.S., White, S.D.M., Efstathiou, G., Davis, M., 1985, {\it Nature}, 
317, 595. 

\pp Frenk, C.S., White, S.D.M., Efstathiou, G., Davis, M., 1988, {\it Ap.J.}, 
327, 507.

\pp Glazebrook, K., Peacock, J.A., Collins, C.A., Miller, L.,  1994,  
\mn \\ 266, 65

\pp Glazebrook, K., Peacock, J.A., Miller, L., Collins, C.A., 1995a,  
\mn \\ 275, 169

\pp Glazebrook, K., Ellis, R., Colless, M., Broadhurst, T., 
    Allington-Smith, J.,  \\ Tanvir, N., 1995b, \mn 273, 157

\pp Glazebrook, K., Ellis, R., Santiago, B., Griffiths, R., 1995c, \\ 
\mn 275, L19

\pp Heyl, J.S., Cole, S., Frenk, C.S., Navarro, J., 1995, \mn 274, 755

\pp Katz, N., Hernquist, L., Weinberg, D.H., 1992, \apj 399, L109

\pp Katz, N., White, S.D.M., 1993,  \apj 412, 455

\pp Kauffmann, G., White, S.D.M., Guiderdoni, B., 1993, \mn  264, 201

\pp Lacey, C.G., Guiderdoni, B., Rocca-Volmerange, B.,  Silk, J., 1993, \apj
  \\ 402,15

\pp Loveday, J., Peterson, B.A., Efstathiou, G., Maddox, S.J., 1992, \apj 390, 338

\pp McGaugh, S., 1994, \nat 367, 538   

\pp Marzke, R.O., Geller, M.J., Huchra, J.P., Corwin, H.G., 1994, 
\aj 108, 437

\pp Mobasher, B., Sharples, R.M., Ellis, 1986, \mn 223, 11

\pp Navarro, J.F.,  White, S.D.M., 1993, \mn  265, 271

\pp Navarro, J.F., Frenk, C.S., White, S.D.M., 
1995, \mn in press

\pp Rees, M.J., Ostriker, J.P., 1977, \mn 179, 541

\pp Silk, J., 1977 \apj 211, 638

\pp Songaila, A., Cowie, L.L., Hu, E.M., Gardner, J.P., 1994, \\
   {\it Ap. J. Suppl., } 94, 461

\pp Steinmetz, M., Muller, E., 1995, \mn 276,  549

\pp Quinn, P.J., Salmon, J.K. and Zurek, W. 1986, {\it Nature}, 322, 
329. 

\pp White, S.D.M., 1995, Les Houches Lecture Notes

\pp White, S.D.M., Rees, M.J. 1978, \mn  183, 341.

\pp Young, P., Lucey, J.L., 1995, in preparation

\end{document}